\documentclass[reprint,showpacs,prl]{revtex4-1}
\usepackage{graphicx}
\usepackage{amssymb}

\begin{document}
\bibliographystyle{aip}

\title{Phonon self-energy corrections to non-zero wavevector phonon modes in single-layer graphene}

\author{P. T. Araujo$^{1,\dag}$, D. L. Mafra$^{1,2,\dag}$, K. Sato$^{3}$, R. Saito$^{3}$, J. Kong$^{1}$ and M. S. Dresselhaus$^{1,4}$}
\affiliation{$^1$ Department of Electrical Engineering and Computer Sciences, Massachusetts Institute of Technology, Cambridge, MA 02139-4307, USA.\\
$^2$ Departamento de F\'{\i}sica, Universidade Federal de Minas Gerais, Belo Horizonte, MG, 30123-970 Brazil.\\
$^3$ Department of Physics, Tohoku University, Sendai 980-8578, Japan.\\
$^4$ Department of Physics, Massachusetts Institute of Technology, Cambridge, MA 02139-4307, USA.\\
$^{\dag}$ These authors contributed equally to this work.}

\date{\today}

\begin{abstract}

Phonon self-energy corrections have mostly been studied theoretically and experimentally for phonon modes with zone-center ($q=0$) wave-vectors. Here, gate-modulated Raman scattering is used to study phonons of a single layer of graphene (1LG) in the frequency range from 2350 to 2750\,cm$^{-1}$, which shows the G$^{\star}$ and the G$^{\prime}$-band features originating from a double-resonant Raman process with $q\neq0$. The observed phonon renormalization effects are different from what is observed for the zone-center $q=0$ case. To explain our experimental findings, we explored the phonon self-energy for the phonons with non-zero wave-vectors ($q\neq0$) in 1LG in which the frequencies and decay widths are expected to behave oppositely to the behavior observed in the corresponding zone-center $q=0$ processes. Within this framework, we resolve the identification of the phonon modes contributing to the G$^{\star}$ Raman feature at 2450\,cm$^{-1}$ to include the iTO+LA combination modes with $q\neq0$ and the 2iTO overtone modes with $q=0$, showing both to be associated with wave-vectors near the high symmetry point $\bf{K}$ in the Brillouin zone.

\end{abstract}

\pacs{73.20.Hb, 73.22.-f, 78.30.Na, 78.67.Ch}
\date{\today.}
\maketitle

Electron-phonon (el-ph) interactions are responsible for many important effects in condensed matter physics \cite{01,02}. In particular, the phonon self-energy, which is mainly due to the el-ph coupling, is a remarkable effect which contributes to both the phonon frequency and decay width renormalizations due to the creation (annihilation) of electron-hole (e-h) pairs through phonon absorption (emission). For non-zero Fermi energy values ($E_{\rm F}\neq0$), phonon self-energy corrections are needed to explain a set of well-known effects, such as the Kohn anomaly \cite{01,03,04,05}, the Peierls transition \cite{01,06,07}, polaron formation \cite{01,08,09}, and other types of phonon renormalizations and perturbations caused by e-h creation (annihilation) due to phonon absorption (emission) \cite{10,11,12,13,24,14}.

Single-layer graphene (1LG) has linear electronic energy dispersions $E(k)$ around the non-equivalent high symmetry points $\bf{K}$ and $\bf{K'}$ in the Brillouin zone as a solution of the Dirac equation which gives massless particle behavior around $\bf{K}$ ($\bf{K'}$) \cite{05,15,16}. However, one cannot properly solve the electronic and vibrational structure for most carbon materials near the Dirac points when considering the adiabatic approximation, which disregards the ionic motion of the carbon ions \cite{15,17,18,19,20}. When the adiabatic approximation cannot be applied \cite{01,02,03,04}, the el-ph interactions must take into account non-adiabatic processes, which give rise to important and strong phonon self-energy corrections \cite{01}. Within second-order perturbation theory, the phonon self-energy can be approximately described as \cite{01,04,24,14}:
\begin{equation}
\Pi(\omega_{\bf{q}}, E_{F})=2\sum_{\bf{kk'}}\frac{|\rm V_{\bf{kk'}}|^2}{\hbar\omega_{\bf{q}}-E^{eh}_{\bf{kk'}}+\it{i}\gamma_{\bf{q}}/2}\times(f_{h}-f_{e})\ \,
\label{eq1}
\end{equation}
where $\bf{k}$ and $\bf{k'}$ are, respectively, wave-vectors for the initial and final electronic states; $\bf{q}\equiv\bf{k-k'}$ is the phonon wave-vector; $E^{eh}_{\bf{kk'}}\equiv(E^{e}_{\bf{k'}}-E^{h}_{\bf{k}})$ is the e-h pair energy; $\omega_{\bf{q}}$ is the phonon frequency; $\gamma_{\bf{q}}$ is the phonon decay width; $f_{h}(f_{e})$ is the Fermi distribution function for holes (electrons) and $\rm V_{\bf{kk'}}$ gives the el-ph coupling matrix element information. For a specific $\omega_{\bf{q}}$, the phonon energy correction due to its self-energy is given by $\hbar\omega_{\bf{q}}-\hbar\omega^{0}_{\bf{q}}={\rm Re}[\Pi(\omega_{\bf{q}}, E_{F})]$, which is the real part of Eq.\ref{eq1}, where $\hbar\omega^{0}_{\bf{q}}$ is the phonon energy in the adiabatic approximation. Likewise, the decay width $\gamma_{\bf{q}}$ is given by the imaginary part ${\rm Im}[\Pi(\omega_{\bf{q}}, E_{F})]$ of Eq.\,\ref{eq1}. Note that while $\gamma_{\bf{q}}/2$, which is self-consistently determined by ${\rm Im}[\Pi(\omega_{\bf{q}}, E_{F})]$ \cite{11,12,13,24}, is always a positive quantity, the quantity $\hbar\omega_{\bf{q}}-\hbar\omega^{0}_{\bf{q}}={\rm Re}[\Pi(\omega_{\bf{q}}, E_{F})]$ is positive if $\hbar\omega_{\bf{q}}>E^{eh}_{\bf{kk'}}$ or negative if $\hbar\omega_{\bf{q}}<E^{eh}_{\bf{kk'}}$. If $\hbar\omega_{\bf{q}}-\hbar\omega^{0}_{\bf{q}}>0$, a phonon hardening in $\omega_{\bf{q}}$ occurs while if $\hbar\omega_{\bf{q}}-\hbar\omega^{0}_{\bf{q}}<0$, a softening occurs.

\begin{figure*}
\includegraphics[angle=0,scale=1]{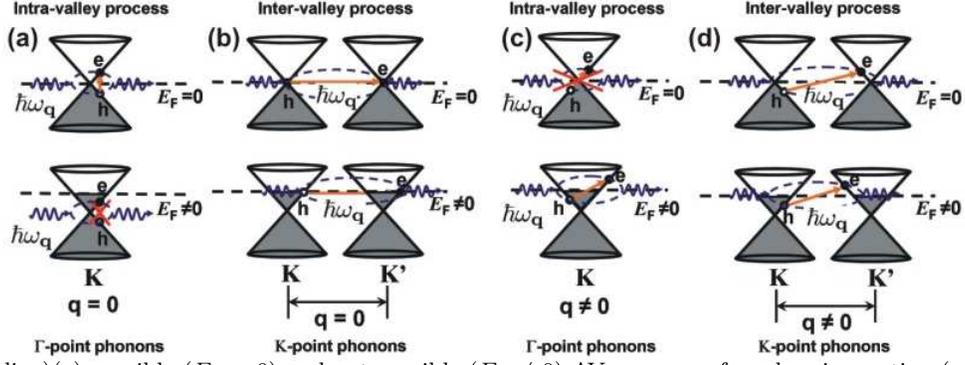}
\vspace{-0.5cm} \caption{(Color online)(a) possible ($E_{\rm F}=0$) and not possible ($E_{\rm F}\neq0$) AV processes for e-h pair creation (annihilation) due to phonon (with energy $\hbar\omega_{\bf{q}}$) absorption (emission). (b) possible ($E_{\rm F}=0$ and $E_{\rm F}\neq0$) $q=0$ (measured from the $\bf{K}$ point) EV processes. (c) not possible ($E_{\rm F}=0$) and possible ($E_{\rm F}\neq0$) AV processes for electron-hole pair creation (annihilation) due to phonon (with energy $\hbar\omega_{\bf{q}}$) absorption (emission) when the phonon wave-vector is not zero ($q\neq0$). (d) possible EV processes for electron-hole pair creation (annihilation) due to phonon (with energy $\hbar\omega_{\bf{q}}$) absorption (emission) when the phonon wave-vector is not zero ($q\neq0$).}
\label{f01}
\end{figure*}

These phonon renormalizations occur any time we have non-zero matrix elements $\rm V_{\bf{kk'}}$ and occupied (unoccupied) initial (final) states, in the sense that an electron-hole pair can be created (annihilated) by a phonon absorption (emission) process as a perturbation. As defined in Fig.\,\ref{f01}, there are two types of electron-phonon interactions, namely intra-valley (AV) (Figs.\,\ref{f01}(a) and (c)) and inter-valley (EV) (Figs.\,\ref{f01}(b) and (d)) processes \cite{x1142}. For an AV process, the initial and final states both occur within the region close to a $\bf{K}$ [$\bf{K'}$] point, while for inter-valley processes, $\bf{K}$ is connected to $\bf{K'}$ (or $\bf{K'}$ to $\bf{K}$), respectively, by a $q\neq0$ phonon. Thus the AV (EV) process corresponds to $\rm {\bf \Gamma}$ (${\bf K}$) point phonons. The phonon wavevector $\bf{q}$ for an AV (EV) process is measured from the $\rm {\bf \Gamma}$ ($\rm {\bf K}$) points and can assume both the $q=0$ or $q\neq0$ conditions. Because of momentum conservation requirements, the $q=0$ condition connects electronic states with the same wave-vector $k$, while the $q\neq0$ condition connects electronic states with different $k$ values.

\begin{figure}
\includegraphics[angle=0,scale=0.7]{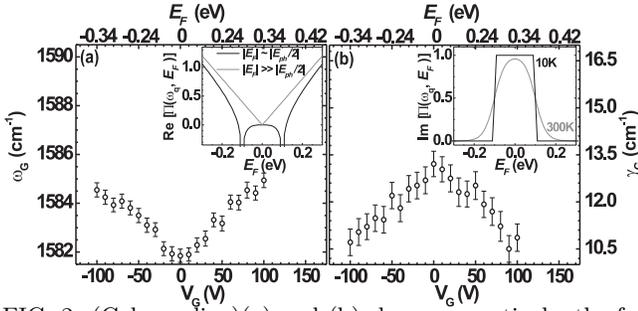}
\vspace{-0.5cm} \caption{(Color online)(a) and (b) show, respectively, the frequency $\omega_{G}$ hardening and decay width $\gamma_{G}$ narrowing for the G-band Raman feature as a function of gate voltage. The insets in (a) and (b) are theoretical pictures of the $E_{\rm F}$ dependence of Re[$\Pi(\omega_{\bf{q}}, E_{F})$] and Im[$\Pi(\omega_{\bf{q}}, E_{F})$] for an AV $q=0$ process. The $E_{F}$ values on the upper scales are obtained with $E_{F}$=$\hbar$$|v_{F}|$$\sqrt{\pi C_{g}V_{G}/e}$, where $C_{g}$=115$aF/\mu$m$^{2}$ is the gate capacitance, $e$ is the electron charge and $|v_{F}|$=1.1$\times$10$^{6}$\,m/s is the Fermi velocity. The Re[$\Pi(\omega_{\bf{q}}, E_{F})$] and Im[$\Pi(\omega_{\bf{q}}, E_{F})$] values in the insets in (a) and (b) were normalized to their maximum values to illustrate the concept of $\omega_{\bf{q}}$ hardening and $\gamma_{\bf{q}}$ narrowing.}
\label{f02}
\end{figure}

\begin{figure*}
\includegraphics[angle=0,scale=0.7]{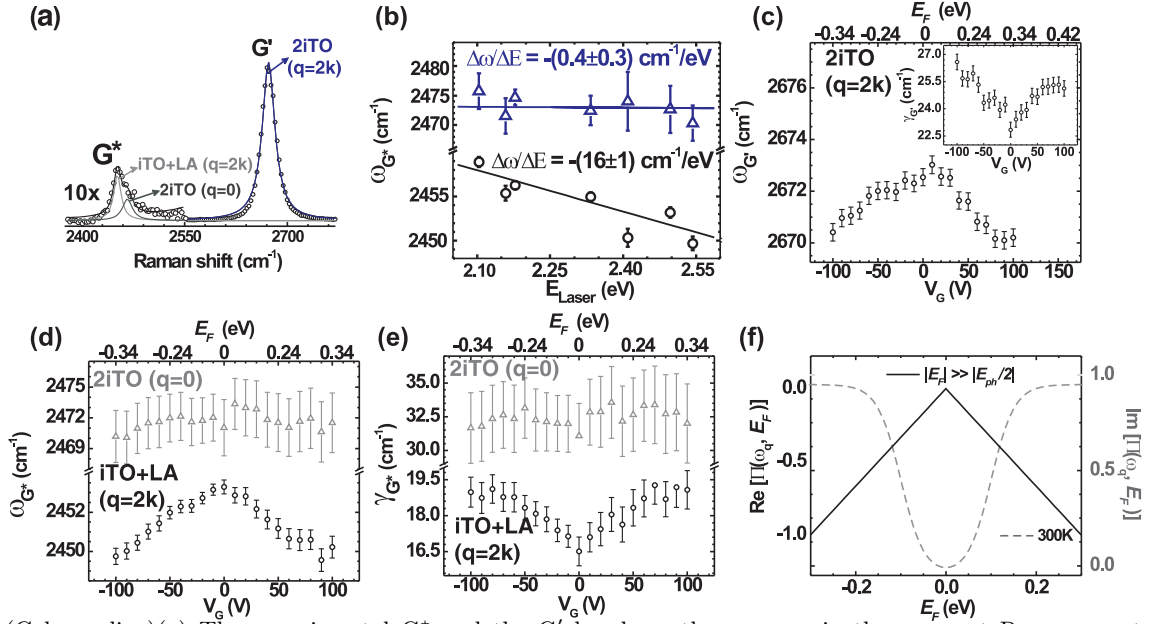}
\vspace{-0.5cm} \caption{(Color online)(a) The experimental G$^{\star}$ and the G$^{\prime}$ bands as they appear in the resonant Raman spectrum. The asymmetric G$^{\star}$ feature is a combination of the iTO+LA ($q=2k$ read from the \textbf{K} point) mode and the 2iTO ($q=0$ read from the \textbf{K} point) mode. The G$^{\prime}$ mode is an overtone of the iTO mode ($q=2k$). For illustrative purposes, the signal of the G$^{\star}$ feature was multiplied by a factor of 10 and the Lorentizian profiles used to fit the spectrum are shown in constructing (a). (b) The frequency dispersion of the G$^{\star}$ peaks as a function of laser energy ($E_{\rm Laser}$) and shows that the iTO+LA ($q=2k$) is a dispersive mode [$-(16\pm1)$\,cm$^{-1}$/eV], while the 2iTO ($q=0$) is non-dispersive [$-(0.4\pm0.3)$\,cm$^{-1}$/eV] \cite{26,25}. (c) The gate voltage $V_{G}$ dependence of the 2iTO ($q=2k$) $\omega_{\bf{G^{\prime}}}$ and $\gamma_{\bf{G^{\prime}}}$ (inset in (c)). (d) and (e) show, respectively, the $\omega_{\bf q}$ and $\gamma_{\bf{q}}$ dependencies on $|E_{\rm F}|$ seen for the iTO+LA and 2iTO modes. (f) shows illustrative predictions for the $V_{G}$-dependence of the phonon frequency correction $\omega_{\bf{q}}-\omega^{0}_{\bf{q}}$ (black solid line) and the corresponding decay width $\gamma_{\bf{q}}$ (grey dashed line) when $q\neq0$, both as a function of $ E_{\rm F}$. In (c), (d) and (e), the $E_{F}$ values were obtained as shown in Fig.\,\ref{f02}. The Re[$\Pi(\omega_{\bf{q}}, E_{F})$] and Im[$\Pi(\omega_{\bf{q}}, E_{F})$] values in (f) were normalized to the maximum value to illustrate the concept of $\omega_{\bf{q}}$ softening and $\gamma_{\bf{q}}$ broadening.}
\label{f03}
\end{figure*}

In the past, most of the discussions of the phonon self-energy renormalizations have been for zone-center phonons ($\bf{\Gamma}$-point) with $q=0$, which can be appreciated by observing the G-band Raman feature evolution in 1LG as the Fermi level is varied (see Fig.\,\ref{f02})\cite{05,10,11,14,21,22,22a}. In the present work, we use gate-modulated resonant Raman spectroscopy (RRS) to address the effect of a gate voltage ($V_{G}$) on the phonon self-energy (Eq.\ref{eq1}), for 1LG systems, in cases where $q\neq0$ (AV and EV processes). These are cases which have not been sufficiently studied previously. Here, we study the double resonance Raman frequency ranges between 2350 and 2850\,cm$^{-1}$, which contains the G$^{\star}$ and the G$^{\prime}$-band spectral features as shown in Fig.\,\ref{f03}(a) \cite{16,25,26}. We show that the phonon renormalization for the $q\neq0$ and $q=0$ phonons measured from the $\bf{K}$ $(\bf{K'})$ points in the Brillouin zone gives a different Fermi-energy dependence to that for the $q=0$ $\bf{\Gamma}$-point phonons. We apply these differences in behavior to show that the G$^{\star}$ feature is composed of two Raman peaks which behave differently from one another as $|V_{G}|$ is varied.

The graphene flakes used in our experiments were obtained by micro-mechanical exfoliation of graphite over silicon substrates with a 300\,nm thick layer of SiO$_{2}$. Next, e-beam lithography was performed to pattern our devices. Then, thermal evaporation of Cr/Au (5 and 80\,nm, respectively) was done. For each $V_{G}$ value, RRS spectra were taken with a 532\,nm wavelength laser source in the backscattering geometry using a 100X objective. The laser power measured from the objective was 1.5\,mW. The spectra were analyzed using Lorentizian line-shapes from which frequencies and decay widths were extracted. Figures\,\ref{f02}(a) and (b) and Figs.\,\ref{f03}(a)-(e) show the experimental results. Note that RRS provides information about both the electronic and vibrational structures, while the $V_{G}$ variation allows for control of $E_{F}$. Due to the difference in behavior between the $q=0$ and $q\neq0$ processes, this combination of techniques provides a precise way to verify the assignments of either overtones and/or a combination of phonon modes. It is worth commenting that the results presented here are observed for several samples and for several locations on the same sample.

The G$^{\star}$ and G$^{\prime}$ features were intentionally chosen for this discussion because: (1) they are the most prominent double-resonance Raman features in the graphene spectrum, offering a convenient platform, together with the G-band feature, to observe experimentally the two different types of phonon renormalizations, one found for the $q=0$ phonons and the other for $q\neq0$, and (2) by using these different phonon renormalization effects, we have solved a long-time discussion in the literature about the G$^{\star}$ feature, showing that the G$^{\star}$ feature is composed of $\underline{\rm both}$ the iTO+LA ($q=2k$) and 2iTO ($q=0$) modes, both measured from the $\bf{K}$-point. In the literature, the G$^{\star}$ feature around 2450\,cm$^{-1}$ has been assigned to either the iTO+LA phonon combination mode ($q=2k$ EV process) \cite{26}, or the 2iTO phonon overtone mode ($q=0$ EV process) \cite{25}, awaiting a more definitive assignment. Note the possibility of two types of double resonance conditions, $q=0$ and $q=2k$, for a phonon with momentum $\bf{q}$ and an electron with momentum $\bf{k}$ \cite{DRspectra,DRspectra2}. We show here that both the 2iTO ($q=0$) and the iTO+LA ($q=2k$) modes are Raman active and therefore both are expected to appear. The G$^{\prime}$ (or 2D) feature at 2670\,cm$^{-1}$ is widely known to be an overtone of the iTO phonon mode ($q=2k$) \cite{16,25,26}. It gives a dispersive phonon frequency as a function of laser energy $E_{\rm Laser}$ which exhibits the value of 103\,cm$^{-1}$/eV \cite{AdoPRB}. The iTO+LA ($q=2k$) combination mode presents a dispersion of -(16$\pm$1)\,cm$^{-1}$/eV (measured in this work), while the 2iTO ($q=0$) overtone mode (also measured in this work) does not disperse (see Fig.\,\ref{f03}(b)). Figure\,\ref{f03}(a) shows that indeed the G$^{\star}$ feature is asymmetric, suggesting that it consists of two Lorentizians peaks rather than just one.

Equation\,\ref{eq1} has previously been explored for the cases where the phonon momentum $\bf{q}$ vanishes ($q=0$) for the AV intra-valley process. In these cases, $\hbar\omega_{\bf{q}}-\hbar\omega^{0}_{\bf{q}}={\rm Re}[\Pi(\omega_{\bf{q}}, E_{F})]$ and $\gamma_{\bf{q}}={\rm Im}[\Pi(\omega_{\bf{q}}, E_{F})]$ give the results illustrated for the renormalization of $\omega_{\bf{q}}$ and $\gamma_{\bf{q}}$, respectively, in the insets of Figs.\,\ref{f02}(a) and (b), respectively. Observing these insets, we conclude that the renormalization is a minimum when $E_{\rm F}=0$. When $|E_{\rm F}|$ becomes approximately half of the phonon energy ($E_{q}/2=\hbar\omega_{\bf{q}}/2$) and greater, the renormalization becomes strong because of decreases in the energy denominator in Eq.\,\ref{eq1} \cite{24}. As an example of phonon renormalization when $q=0$ for the AV process (see Fig.\,\ref{f01}(a)), the $\omega_{G}$ and $\gamma_{G}$ variations of the G-band Raman feature are shown in Figs.\,\ref{f02}(a) and (b), respectively, as $V_{G}$ is varied. The G-band feature corresponds to the first order $q=0$ iTO (in-plane transverse optical) and iLO (in-plane longitudinal optical) phonon branches around the $\bf{\Gamma}$ point. As regards the $\omega_{G}$ and $\gamma_{G}$ behaviors as a function of $V_{G}$, the experimental results (Figs.\,\ref{f02}(a) and (b)) are in good agreement with literature predictions \cite{04,05,21,22,22a}, which show a $\omega_{G}$ hardening and $\gamma_{G}$ narrowing when $V_{G}$ increases. Next, we report the new experimental results for phonons corresponding to the cases $q=0$ EV (inter-valley) and $q\neq0$ AV/EV.

Both the G$^{\star}$ iTO+LA mode at 2450-53\,cm$^{-1}$ and the G$^{\prime}$ mode at 2670-73\,cm$^{-1}$ are EV double-resonance Raman processes with $q\neq0$ (see Fig.\,\ref{f03}(a)) and, as shown in Figs.\,\ref{f03}(c)-(e), they show a different behavior when $V_{\rm G}$ increases compared to the behavior observed for the AV $q=0$ process. Starting with the G$^{\prime}$-band feature (2iTO EV process with $q=2k$ measured from the $\bf{K}$-point), it is seen that its frequency $\omega_{\rm G^{\prime}}$ decreases with increasing $|V_{G}|$ (Fig.\,\ref{f03}(c)), while its decay width $\gamma_{\rm G'}$ increases with increasing $|V_{G}|$ (see the inset in Fig.\,\ref{f03}(c)). Here, we see that the same behavior is observed for the iTO+LA mode frequency $\omega_{\rm iTO+LA}$ (EV process with $q=2k$ measured from the $\bf{K}$-point), as shown in Fig.\,\ref{f03}(d), and for its decay width $\gamma_{\rm iTO+LA}$, as shown in Fig.\,\ref{f03}(e). The 2iTO G$^{\star}$ feature at 2470-73\,cm$^{-1}$, which is a $q=0$ EV process around the ${\bf K}$ point (see Fig.\,\ref{f01}(b)), is observed in Figs.\,\ref{f03}(d) and (e) to show a frequency $\omega_{\rm 2iTO}$ and decay width $\gamma_{\rm 2iTO}$ that almost do not change with increasing $|V_{G}|$. This behavior shows that the 2iTO ($q=0$) mode \cite{24} couples weakly to the electronic states in graphene and therefore its phonon self-energy corrections are negligible. These findings show a different $\omega_{\bf{q}}$ and $\gamma_{\bf{q}}$ dependence with $|E_{\rm F}|$ when compared to the AV $q=0$ phonon process (see Fig.\,\ref{f02}).

To explain our experimental findings, a phenomenological formulation for the phonon self-energy for the EV $q=0$ and AV/EV $q\neq0$ processes in single-layer graphene is presented as follows. In the case of AV processes for the $q=0$ phonons (Fig.\,\ref{f01}(a)), which applies for the G-band feature (see Fig.\,\ref{f02}(a) and (a)), the renormalizations are a minimum when $E_{\rm F}=0$, which implies a phonon frequency ($\omega_{\bf{q}}$) softening and phonon decay width ($\gamma$) broadening, and both $\omega_{\bf{q}}$ hardening and $\gamma_{\bf{q}}$ narrowing with increasing $|E_{\rm F}|$, as shown in the insets in Figs.\,\ref{f02}(a) and (b). More precisely, the phonon renormalization shows its smallest values when $E_{F}=\pm\hbar\omega_{\bf{q}}/2$, which represent two singularities in Eq.\,\ref{eq1}, as shown by the black solid curve in the inset of Fig.\,\ref{f02}(a). These singularities give rise to what is commonly known as Kohn-anomalies. However, it is difficult to experimentally observe these Kohn-anomalies near room temperature because of thermal excitations (relaxations). The absence of such divergencies is also attributed to a large non-uniformity in the density of carriers due to foreign molecules and charge traps in the substrate \cite{22a}. Therefore, the condition $|E_{F}|>>\hbar\omega_{\bf q}/2$ usually applies and the gray solid line-shape in the inset of Fig.\,\ref{f02}(a) is expected to be observed, which means that $\hbar\omega_{\bf{q}}-\hbar\omega^{0}_{\bf{q}}$ is always positive when $|E_{\rm F}|$ increases \cite{13,24}. Indeed, for $q=0$ the $E^{eh}_{\bf{kk'}}$ must be larger than 2$|E_{\rm F}|$, and if 2$|E_{\rm F}|>>\hbar\omega_{\bf{q}}$, no real e-h pair due to phonon absorption(emission) will be created and $\hbar\omega_{\bf{q}}-\hbar\omega^{0}_{\bf{q}}$ is positive. For simplicity, the discussion hereafter is based on the assumption that $|E_{F}|>>\hbar\omega_{\bf q}/2$ for non-zero $|E_{\rm F}|$.

Note that, a different behavior is expected for the $q=0$ phonon (measured from $\bf{K}$-point) in the EV process shown in Fig.\,\ref{f01}(b), which explains the G$^{\star}$ 2iTO mode behavior as $|V_{G}|$ is varied. The probability that an electron-hole pair exists at $E_{\rm F}=0$ (upper line of Fig.\,\ref{f01}) is small since the density of states at $E_{F}$ almost vanishes and this implies that no $E_{F}$-dependence for the frequency $\omega_{\bf{q}}$ and width $\gamma_{\bf{q}}$ is expected. When $|E_{\rm F}|$ increases (lower line in Fig.\,\ref{f01}), the probability for a $q=0$ phonon to connect equal $\bf{k}$ and $\bf{k'}$ states ($k=k'$) increases, which means that the number of e-h pairs increases and thus the renormalization effect could be appreciable with increasing $|E_{\rm F}|$. However, for $q=0$ EV processes, the $E^{eh}_{\bf{kk'}}$ will be always approximately null ($E^{eh}_{\bf{kk'}}\sim0$). This means that $\omega_{\bf{q}}-\omega^{0}_{\bf{q}}$ will be a small correction and, therefore, small $\omega_{\bf{q}}$ and $\gamma_{\bf{q}}$ renormalizations are expected for any $|E_{\rm F}|$ value (weak $E_{\rm F}$-dependence).

By considering phonon modes with $q\neq0$ (AV and EV processes) as shown in Figs.\,\ref{f01}(c) and (d), the phonon wave-vectors are either around the $\bf{\Gamma}$ point or around the $\bf{K}$ point. This case explains the G$^{\prime}$ and the G$^{\star}$ iTO+LA mode behaviors as $|V_{G}|$ is varied. Since the phonon energy dispersion for graphene has a much smaller slope than that for the electronic energy dispersions \cite{16}, there is essentially no coupling between $q\neq0$ phonons with e-h pairs if $E_{\rm F}=0$ and therefore phonon renormalizations in this case do not take place in a resonant way, {\it i.e.} where $E_{\bf{k^{\prime}}}-E_{\bf{k}}=\hbar\omega_{\bf q}$. This means that no phonons with $q\neq0$ can connect electronic states with different $\bf{k}$ and $\bf{k'}$ at $E_{\rm F}=0$ and as a result, the matrix elements $V_{\bf kk^{\prime}}$ in Eq.\,\ref{eq1} are close to zero and essentially no self-energy corrections occur. Precisely speaking, in the case of the EV process (Fig.\,\ref{f01}(d)), the scattering is possible for $E_{\rm F}=0$, but the density of states (DOS) is very small. However, when $E_{\rm F}\neq0$, phonon modes with $q\neq0$ can now connect electronic states with different {\bf k} and {\bf k$^{\prime}$}, giving rise to a strong electron-phonon coupling and then the phonon self-energy corrections become important (see Figs.\,\ref{f01}(c) and (d)). Such a $q\neq0$ phonon will be observed in the double resonance Raman spectra \cite{DRspectra}, and in this case one can observe defect-induced features, overtones or combination modes \cite{05,14,16}.

For $q\neq0$ AV and EV processes, the requirement $E^{eh}_{\bf{kk^{\prime}}}>2|E_{\rm F}|$ no longer exists and for most of the situations the condition $E^{eh}_{\bf{kk^{\prime}}}>\hbar\omega_{\bf{q}}$ will be satisfied and then $\omega_{\bf{q}}-\omega^{0}_{\bf{q}}$ will be negative. Now for the $q\neq0$ AV and EV cases, only phonon softening and decay width broadening are predicted to occur. This requires that $\omega_{\bf{q}}-\omega^{0}_{\bf{q}}$ and $\gamma_{\bf{q}}$ must behave oppositely to the behavior observed in the insets of Fig.\,\ref{f02}. This opposite behavior is illustrated in Fig.\,\ref{f03}(f), where it is seen that the frequency correction $\omega_{\bf{q}}-\omega^{0}_{\bf{q}}$ (black solid line) must become more negative with increasing $|E_{\rm F}|$ while the decay width (grey dashed line) must increase with increasing  $|E_{\rm F}|$.

In summary, the widely studied intra-valley AV $q=0$ case \cite{04,05,21,22,22a} shows that when $E_{\rm F}=0$ the phonon renormalization is a minimum. The phonon frequency then shows a hardening while the decay width shows a narrowing when $|E_{\rm F}|$ increases. Here, we have studied the phonon self-energy correction for phonon modes with $q\neq0$ (AV and inter-valley EV processes) and $q=0$ (EV process) from a theoretical and experimental point of view. In the $q\neq0$ cases (oppositely to what is observed for the $q=0$ AV process), the phonon renormalization is a maximum when $E_{\rm F}=0$ and the phonon frequency softens while the decay width broadens with increasing $|E_{\rm F}|$. Namely, while the decay width $\gamma_{\bf{q}}$ is always positive, the frequency correction $\omega_{\bf{q}}-\omega^{0}_{\bf{q}}$ is positive for the $q=0$ AV case but negative for the $q\neq0$ AV/EV cases. For the EV $q=0$ case, $E^{eh}_{\bf{kk^{\prime}}}\sim0$ and a weak and small $\omega_{\bf{q}}$ and $\gamma_{\bf{q}}$ dependence with $E_{\rm F}$ is expected. In this context, gate-modulated resonant Raman spectroscopy of overtones and of a combination of phonon modes provides a powerful technique to assign the phonons participating in the formation of each overtone or combination mode, to identify whether a Raman feature is associated with the $q=0$ or $q\neq0$ processes and to determine how a given phonon mode is coupled to the electronic states of single-layer graphene. As shown in Figs.\,\ref{f03}(a)-(f), we applied these combined techniques to study the G$^{\star}$ and G$^{\prime}$ modes, which are the most prominent double resonance $q\neq0$ Raman features in the graphene spectrum. Our theoretical approach satisfactorily explains the experimental results and within this framework, we also showed that the G$^{\star}$ mode is an asymmetric peak composed by $\underline{\rm both}$, the iTO+LA combination mode, which is an EV $q=2k$ process with a $\omega_{\bf{q}}-\omega^{0}_{\bf{q}}<0$ renormalization and the 2iTO overtone mode, which is an EV $q=0$ process with a weak phonon renormalization, thereby resolving a long-time discussion in the literature.

\section*{Acknowledgments}

P.T.A. and D.L.M. acknowledge CNPq and NSF-DMR 10-04147. R.S. acknowledges MEXT grant (No.20241023). M.S.D acknowledges NSF-DMR 10-04147.

%\bibliography{/rsaito/bib/mgm,/rsaito/bib/gic,/rsaito/bib/mag_gic,/rsaito/bib/c60,/rsaito/bib/carbon,/rsaito/bib/fiber,/rsaito/bib/alex,/rsaito/bib/saito,ado08-KP}

% \end{document}

\end{document}